\pdfoutput=1
\RequirePackage{ifpdf}
\ifpdf 
\documentclass[pdftex]{sigma}
\else
\documentclass{sigma}
\fi

\numberwithin{equation}{section}

\newtheorem{Theorem}{Theorem}[section]
\newtheorem{Proposition}[Theorem]{Proposition}
 { \theoremstyle{definition}
\newtheorem{Definition}[Theorem]{Definition}
\newtheorem{Remark}[Theorem]{Remark} }

\usepackage{tikz,mathdots}

\newcommand{\p}{\partial}
\newcommand{\bdt}{{\bf t}}

\newcommand{\g}{\mathfrak{g}}
\newcommand{\fb}{\mathfrak{b}}
\newcommand{\fn}{\mathfrak{n}}

\newcommand{\h}{\mathfrak{h}}

\newcommand{\CC}{\mathbb{C}}
\def\L{\mathcal L}

\def\sl{{\rm sl}}
\def\H{{\cal H}}
\newcommand{\ZZ}{\mathbb{Z}}
\def\bdx{{\bf x}}

\newcommand{\Ker}{\operatorname{Ker}}
\newcommand{\ad}{\operatorname{ad}}

\usepackage{array}
\newcolumntype{M}[1]{>{\centering\arraybackslash}m{#1}}
\newcolumntype{N}{@{}m{0pt}@{}}

\begin{document}

\allowdisplaybreaks

\newcommand{\arXivNumber}{1709.07309}

\renewcommand{\PaperNumber}{104}

\FirstPageHeading

\ShortArticleName{Drinfeld--Sokolov Hierarchies, Tau Functions, and Generalized Schur Polynomials}

\ArticleName{Drinfeld--Sokolov Hierarchies, Tau Functions,\\ and Generalized Schur Polynomials}

\Author{Mattia CAFASSO~$^{\dag}$, Ann DU CREST DE VILLENEUVE~$^{\dag}$ and Di YANG~$^{\ddag\S}$}

\AuthorNameForHeading{M.~Cafasso, A.~du Crest de Villeneuve and D.~Yang}

\Address{$^{\dag}$~LAREMA, Universit\'e d'Angers, 2 boulevard Lavoisier, Angers 49000, France}
\EmailD{\href{mailto:cafasso@math.univ-angers.fr}{cafasso@math.univ-angers.fr}, \href{mailto:ducrest@math.univ-angers.fr}{ducrest@math.univ-angers.fr}}

\Address{$^{\ddag}$~Max Planck Institute for Mathematics, Vivatsgasse 7, Bonn 53111, Germany}
\EmailD{\href{mailto:diyang@mpim-bonn.mpg.de}{diyang@mpim-bonn.mpg.de}}
\Address{$^{\S}$~School of Mathematical Sciences, University of Science and Technology of China,\\
\hphantom{$^{\S}$}~Hefei 230026, P.R.~China}
\EmailD{\href{mailto:diyang@ustc.edu.cn}{diyang@ustc.edu.cn}}

\ArticleDates{Received April 28, 2018, in final form September 19, 2018; Published online September 27, 2018}

\Abstract{For a simple Lie algebra $\mathfrak{g}$ and an irreducible faithful representation $\pi$ of $\mathfrak{g}$, we introduce the Schur polynomials of $(\mathfrak{g},\pi)$-type. We then derive the Sato--Zhou type formula for tau functions of the Drinfeld--Sokolov (DS) hierarchy of $\mathfrak{g}$-type. Namely, we show that the tau functions are linear combinations of the Schur polynomials of $(\mathfrak{g},\pi)$-type with the coefficients being the Pl\"ucker coordinates. As an application, we provide a way of computing polynomial tau functions for the DS hierarchy. For $\mathfrak{g}$ of low rank, we give several examples of polynomial tau functions, and use them to detect bilinear equations for the DS hierarchy.}


\Classification{37K10; 17B80}

\section{Introduction}

Given a simple Lie algebra $\g$ over $\mathbb{C}$, Drinfeld and Sokolov in \cite{DS} explained how to associate to it a family of commuting bi-Hamiltonian PDEs known as the Drinfeld--Sokolov hierarchy of $\g$-type. Nowadays, Drinfeld--Sokolov (DS) hierarchies are certainly among the most studied examples of integrable systems; one of their remarkable properties is that they are tau-symmetric \cite{BDY3, DLZ,DZ-norm, Wu}, meaning that they admit the so-called tau function of an arbitrary solution to the hierarchy. For the case $\g= \sl_{n+1}(\CC)$ the DS hierarchy of $\g$-type coincides (under a particular choice of the DS gauge \cite{BFRFW, DS}) with the Gelfand--Dickey hierarchy, and so, in particular, for $n = 1$, with the celebrated Korteweg--de Vries (KdV) hierarchy. It is known that tau functions of the Gelfand--Dickey hierarchies can be expressed as linear combinations of Schur polynomials with the coefficients being Pl\"ucker coordinates \cite{Dickey,Macdonald, Sato}. In this short paper we aim to generalize this fact and its development in \cite{BaY,Zhou} to an arbitrary given Lie algebra $\g$. The generalization will depend on matrix realizations of $\g$ (note that the tau function itself is independent of the realizations of $\g$ \cite{BDY3}!). Indeed, one of our main observations is that the generalization of Schur polynomials should be associated with a faithful representation.

As an application of our result, we describe a systematic way of finding {\it simple solutions} (here a~simple solution means solution whose tau function is a polynomial or a fractional power of a~polynomial) of the DS hierarchy of $\g$-type. Of course, in the case of the hierarchies of type~$A_n$, we recover the well-known results; actually, polynomial tau functions of these hierarchies (more generally of the Kadomtsev--Petviashvili (KP) hierarchy) had been studied for many years, due to their relations with B\"acklund transformations \cite{AdMPoly} and the dynamical systems of Calogero type (see for instance~\cite{WilsonAdelic} and the references therein). Moreover, it had been proved that the polynomial tau functions of the so-called BKP hierarchy can be written in terms of the projective representations of the symmetric group~\cite{You}, and the BKP hierarchy, moreover, contains as reductions some of the DS hierarchies of $D_n$-type, as explained in~\cite{DJKM1}. Nevertheless, it seems to us that a~syste\-ma\-tic approach to the study of polynomial tau functions associated to the general case (i.e., for an arbitrary Lie algebra) is still missing, and this paper gives a first result in this direction. The polynomial tau functions we obtain are, actually, quite non-trivial, and can also be used to give some explicit information about the structure of the bilinear equations for the hierarchy.

In order to state precisely our results, we need to fix some notations about finite-dimensional Lie algebras~\cite{Cartan,Kostant}, loop algebras~\cite{DS,Kac} and Toeplitz determinants~\cite{Cafasso}. Let $\g$ be a simple Lie algebra over $\mathbb{C}$ of rank~$n$, and $h$, $h^{\vee}$ the Coxeter and dual Coxeter numbers, respectively. Fix~$\h$ a Cartan subalgebra of~$\g$. Take $\Pi=\{\alpha_1,\dots,\alpha_n\}\subset\h^*$ a set of simple roots, and let $\triangle\subset \h^*$ be the root system. We know that $\g$ has the root space decomposition
\begin{gather*}
\g=\h\oplus \bigoplus_{\alpha\in\triangle} \g_{\alpha} .
\end{gather*}
Let $\theta$ denote the highest root with respect to $\Pi$, and $(\cdot|\cdot)\colon \g \times\g\rightarrow \mathbb{C}$ the {\it normalized} Cartan--Killing form, i.e., $(\theta|\theta)=2$. For a root $\alpha\in \triangle$, denote by~$H_{\alpha}$ the unique vector in $\h$ satisfying $(H_\alpha|H_\beta)=(\alpha|\beta)$, $\forall\, \beta\in \triangle$.

Let $E_i\in\g_{\alpha_i}, F_i\in\g_{-\alpha_i}$, $H_i={2H_{\alpha_i}}/{(\alpha_i|\alpha_i)}$ be a set of Weyl generators of~$\g$. They satisfy
\begin{gather*}
[E_i,F_i]=H_i , \qquad [H_i,E_j]=A_{ij} E_j ,\qquad [H_i,F_j]=- A_{ij} F_j ,\qquad 1\leq i,j\leq n ,
\end{gather*}
where $\bigl(A_{ij}\bigr)_{i,j = 1}^n$ is the Cartan matrix of $\g$. Choose $E_{-\theta}\in \g_{-\theta}$, $E_\theta \in\g_\theta$, normalized by the conditions $(E_\theta | E_{-\theta})=1$ and $\omega(E_{-\theta})=-E_\theta$, where $\omega\colon \g\to\g$ is the Chevalley involution. Let $I_+:=\sum\limits_{i=1}^n E_i$ be a principal nilpotent element of~$\g$. Denote by $L(\g)=\g \otimes \CC\big[\lambda, \lambda^{-1}\big]$ the loop algebra of~$\g$. On $L(\g)$ there is the {\it principal gradation} defined by assigning
\begin{gather*}
\deg E_i=1 , \qquad \deg H_i=0 , \qquad \deg F_i=-1 , \qquad i=1, \dots, n , \qquad \deg\lambda=h,
\end{gather*}
such that $L(\g)$ decomposes into homogeneous subspaces
\begin{gather*}
L(\g)=\bigoplus_{j\in \mathbb{Z}} L(\g)^j .
\end{gather*}
Here, elements in $L(\g)^j$ have degree $j$.
Define $\Lambda \in L(\g)$ by
\begin{gather*} 
\Lambda=I_+ +\lambda E_{-\theta} .
\end{gather*}
Clearly, $\Lambda$ is homogeneous of degree $1$. Denote by $L(\g)^{<0}$ the set of elements in $L(\g)$ with negative degrees, similarly, by~$L(\g)^{\leq 0}$ elements with non-positive degrees.

It was shown in \cite{Kac1978,Kostant} that $\Ker \ad_\Lambda\subset L(\g)$ has the following decomposition
\begin{gather*}
\Ker \ad_\Lambda=\bigoplus_{\ell \in E} \mathbb{C} \Lambda_\ell ,\qquad \deg\Lambda_\ell=\ell \in E:=\bigsqcup_{i=1}^n (m_i+h \mathbb{Z}) ,
\end{gather*}
where the integers $m_1, \dots,m_n$ are the exponents of $\g$, and $E$ is called the set of exponents of~$L(\g)$. We use $E_+$ to denote the set of positive exponents. The elements $\Lambda_i$ commute pairwise
\begin{gather*}
[\Lambda_i, \Lambda_j ]=0 ,\qquad \forall\, i,j\in E .
\end{gather*}
They can be normalized by
\begin{gather*}
 \Lambda_{m_a+kh}=\Lambda_{m_a} \lambda^k ,\qquad k\in \mathbb{Z} ,\qquad (\Lambda_{m_a}| \Lambda_{m_b})=h \lambda \delta_{a+b, n+1} .
\end{gather*}
In particular, we can choose $\Lambda_1=\Lambda$.

Let us now take
\begin{gather}\label{repi}
\pi\colon \ \g \rightarrow {\rm gl}(m,\mathbb{C})
\end{gather}
an irreducible faithful representation. When the representation is fixed and no confusion can arise, we sometimes write $\pi(b)$ simply as $b$ for $b\in \g$, and write $\pi(\g)$ simply as~$\g$. Our gene\-ra\-lization will be based on the infinite Grassmannian approach \cite{Sato,SW} and the related Pl\"ucker coordinates.

\medskip

\noindent
\textbf{Notations.} a) For $M = \sum\limits_{k\in \mathbb{Z}} M_k \lambda^k$ with $M_k\in{\rm gl}(m,\mathbb{C})$,
define the {\it Laurent matrix} $\mathbb{L}(M)$ associated with $M$ by
\begin{gather*}
\bigl[\mathbb{L}(M)\bigr]_{IJ}=M_{I-J} , \qquad I,J \in\mathbb{Z} ,
\end{gather*}
as in Fig.~\ref{MVC0}. Here and below, we use the capital-letter indices $I,J,K,\dots$ for block row/column coordinates. We note that the rows and columns of $\mathbb{L}(M)$ are labeled by integers and that we divide this matrix into blocks of size $m \times m$. More precisely, the entries of $\mathbb{L}(M)$ satisfy $\mathbb{L}(M)_{Q_1 m+p_1, Q_2 m + p_2} = (M_{Q_1-Q_2})_{p_1+1,p_2+1}$ for all $Q_1,Q_2\in \ZZ$ and
$p_1,p_2=0,\dots,m-1$.

\begin{figure}[htb!]\centering
\begin{tikzpicture}[scale=0.5]
\draw[very thick] (-4,0)--(4,0);
\draw[very thick] (0,-4)--(0,4);
\draw (-4,2) rectangle (0,4);
\draw (-4,0) rectangle (0,2);
\draw (-4,-2) rectangle (0,0);
\draw (-4,-4) rectangle (0,-2);
\draw (-2,2) rectangle (0,4);
\draw (-2,0) rectangle (0,2);
\draw (-2,-2) rectangle (0,0);
\draw (-2,-4) rectangle (0,-2);
\draw (0,2) rectangle (2,4);
\draw (0,0) rectangle (2,2);
\draw (0,-2) rectangle (2,0);
\draw (0,-4) rectangle (2,-2);
\node at (-5,5) {$\ddots$};
\node at (-3,5) {$\vdots$};
\node at (-1,5) {$\vdots$};
\node at (1,5) {$\vdots$};
\node at (3,5) {$\vdots$};
\node at (-3,-5) {$\vdots$};
\node at (-1,-5) {$\vdots$};
\node at (1,-5) {$\vdots$};
\node at (3,-5) {$\vdots$};
\node at (5,5) {$\iddots$};
\node at (5,3) {$\cdots$};
\node at (5,1) {$\cdots$};
\node at (5,-1) {$\cdots$};
\node at (5,-3) {$\cdots$};
\node at (-5,3) {$\cdots$};
\node at (-5,1) {$\cdots$};
\node at (-5,-1) {$\cdots$};
\node at (-5,-3) {$\cdots$};
\node at (-5.25,-4.75) {$\iddots$};
\node at (5.25,-4.75) {$\ddots$};
\node at (1,3) {$M_{-2}$};
\node at (1,1) {$M_{-1}$};
\node at (1,-1) {$M_0$};
\draw (2,2) rectangle (4,4);
\draw (2,0) rectangle (4,2);
\draw (2,-2) rectangle (4,0);
\draw (2,-4) rectangle (4,-2);
\node at (3,3) {$M_{-3}$};
\node at (3,1) {$M_{-2}$};
\node at (3,-1) {$M_{-1}$};
\node at (3,-3) {$M_0$};
\node at (1,3) {$M_{-2}$};
\node at (1,1) {$M_{-1}$};
\node at (1,-1) {$M_0$};
\node at (1,-3) {$M_{1}$};
\node at (-1,3) {$M_{-1}$};
\node at (-1,1) {$M_0$};
\node at (-1,-1) {$M_{1}$};
\node at (-1,-3) {$M_{2}$};
\node at (-3,3) {$M_0$};
\node at (-3,1) {$M_{1}$};
\node at (-3,-1) {$M_{2}$};
\node at (-3,-3) {$M_{3}$};
\end{tikzpicture}
\caption{The Laurent matrix $\mathbb{L}(M)$.}\label{MVC0}
\end{figure}

b) $\mathbb{Y}$ will denote the set of all partitions; for $\nu = (\nu_1\geq \nu_2 \geq\cdots)\in \mathbb{Y}$, denote by $\ell(\nu)$ the {\it length} of $\nu$, by $|\nu|$ the {\it weight} of $\nu$, i.e., $\ell(\nu)$ is the number of non-zero components of $\nu$ and $|\nu|=\nu_1+\dots+\nu_{\ell(\nu)}$. Also, we denote by $\nu=\bigl( k_1,\dots,k_{d(\nu)} | l_1,\dots, l_{d(\nu)} \bigr)$ the Frobenius notation of~$\nu$, for which we recall briefly as follows. First, $d(\nu)$ is the number of squares in the main diagonal of the Young diagram realization of~$\nu$ (these squares have coordinates $(i,i)$, $i=1,\dots,d(\nu)$). The integer $k_i$ is the number of squares in the same row strictly to the right of the square~$(i,i)$, and $l_i$ is the number of squares in the same column strictly below the square~$(i,i)$. For example, the partition $\nu = (5441)$ has length $\ell(\nu) = 4$, weight $|\nu| = 14$, and can be written in the Frobenius notation as $\nu = (421|310)$, as illustrated below.
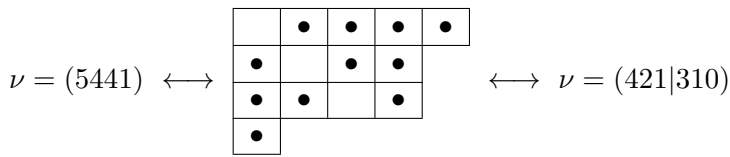
\begin{gather*}
\nu = (5441) \ \longleftrightarrow \
\begin{tabular}{|c|c|c|c|c|c}
\hline
 & $\bullet$ & $\bullet$ & $\bullet$ & $\bullet$\\
\hline
$\bullet$ & & $\bullet$ & $\bullet$\\
\cline{1-4}
$\bullet$ & $\bullet$ & & $\bullet$\\
\cline{1-4}
$\bullet$\\
\cline{1-1}
\end{tabular}
\ \longleftrightarrow \ \nu = (421|310)
\end{gather*}

\begin{Definition}Let $ \xi := \sum\limits_{ \ell \in E_+} t_\ell \Lambda_\ell $ with $t_\ell$, $\ell\in E_+$, being indeterminates, and let $s$ denote the Laurent matrix associated with $e^\xi$, namely,
\begin{gather}
s := \mathbb{L} \bigl(e^{\xi}\bigr) . \label{defs}
\end{gather}
The Schur polynomials of $(\g,\pi)$-type are labelled by partitions and defined by
\begin{gather*}
 s_\nu := \det (s_{i-1, j-\nu_j-1} )_{i,j=1}^{\ell(\nu)} ,\qquad \nu \in \mathbb{Y}-\varnothing , \qquad s_{\varnothing}:=1 .
\end{gather*}
\end{Definition}

\begin{Definition}In the case $\pi$ is taken as the adjoint representation of~$\g$, we call $s_\nu$,~$\nu\in \mathbb{Y}$, the intrinsic Schur polynomials of $\g$-type.
\end{Definition}

\begin{Remark}\label{1.3}In the case $\g=A_n$, take $\pi(\g)$ the well-known matrix realization of~$\g$, i.e., $\pi(\g)= \sl_{n+1}(\mathbb{C})$. We have $\Lambda= \sum\limits_{i=1}^n E_{i,i-1} + \lambda E_{1,n+1}$, where $E_{i,j}$ denotes the $(n+1)\times(n+1)$ matrix with~1 at the intersection of row~$i$ and column~$j$, and~0 elsewhere. The Schur polynomials of $(\g,\pi)$-type then coincide with the Schur polynomials \cite{Macdonald} under the restriction $t_{(n+1)k}\equiv 0$, $k=1,2,3,\dots$.
\end{Remark}

\begin{Definition} For any $X\in \lambda^{-1}\g\bigl[\bigl[\lambda^{-1}\bigr]\bigr]$, denote by $r_X$ the Laurent matrix associated with~$e^X$, that is
\begin{gather}\label{defr}
r_X:=\mathbb{L}\bigl(e^X\bigr) .
\end{gather}
For $\nu=(\nu_1,\dots,\nu_{\ell(\nu)})\in \mathbb{Y}$, define
\begin{gather*}
r_{X,\nu} := \det \left(r_{X,i-\nu_i-1, j-1}\right)_{i,j=1}^{\ell(\nu)} .
\end{gather*}
\end{Definition}

\begin{Definition} For $\xi = \sum\limits_{ \ell \in\mathbb{E}_+} t_\ell \Lambda_\ell$ (as above), and for any $X\in \lambda^{-1}\g\bigl[\bigl[\lambda^{-1}\bigr]\bigr]$, define mat\-ri\-ces~$D_{IJ}$ and $Z_{X,IJ}$ ($I,J\geq 0$) by
\begin{gather}
 \frac{I-e^{\xi(\lambda)} e^{-\xi(\mu)}}{\lambda-\mu} = \sum_{I,J=0}^\infty D_{IJ} \lambda^{I+1} \mu^{J+1} , \label{ms} \\
 \frac{I-e^{X(\lambda)} e^{-X(\mu)}}{\lambda-\mu} = \sum_{I,J=0}^\infty Z_{X, IJ} \lambda^{-I-1} \mu^{-J-1} . \label{mac}
\end{gather}
Define $s_{(i|j)}$, $r_{X,(i|j)}$, $i,j\geq 0$, via
\begin{gather*}
 (D_{IJ})_{ab} = s_{(m \cdot I+a-1| m\cdot J+m-b)} ,\qquad (Z_{X, IJ})_{ab} = r_{X,(m\cdot I+m-a| m\cdot J+b-1)} ,
\end{gather*}
where $a,b=1,\dots,m$. We call $Z_{X, IJ}$ the matrix-valued affine coordinates and $r_{X, (i|j)}$ the affine coordinates.
\end{Definition}

\begin{Remark}The matrix-valued affine coordinates $Z_{X, IJ}$ and their generating formula~\eqref{mac} were introduced in~\cite{BaY} by F.~Balogh and one of the authors of the present paper for the $\sl_2(\CC)$ case.
\end{Remark}

The following theorem is the main result of the paper. Denote by~$\kappa$ the constant such that
\begin{gather}\label{defkappa}
	(a|b) = \kappa \operatorname{Tr} (\pi(a)\pi(b)) , \qquad \forall\, a,b\in\g .
\end{gather}

\begin{Theorem} \label{main}
For any $X\in \lambda^{-1}\g\bigl[\bigl[\lambda^{-1}\bigr]\bigr]$, the formal series~$\tau$ defined by
\begin{gather} \label{SZ}
\tau := \biggl(\sum_{\nu \in \mathbb{Y}} r_{X,\nu} s_\nu \biggr)^\kappa
\end{gather}
is a tau function of the Drinfeld--Sokolov hierarchy of $\g$-type.
Moreover, $s_\nu$ and $r_{X,\nu}$ have the following expressions
\begin{gather}
 s_\nu = \det \big(s_{(k_i|l_j)}\big)_{i,j=1}^{d(\nu)} , \label{sz1}\\
 r_{X,\nu} = (-1)^{l_1+\dots+l_{d(\nu)}}\det \big(r_{X,(k_i|l_j)}\big)_{i,j=1}^{d(\nu)} . \label{sz2}
\end{gather}
\end{Theorem}

We refer to \eqref{SZ}--\eqref{sz2} as the Sato--Zhou type formula for tau functions of the DS hierarchy.

\begin{Remark}As the reader might already have noticed, here the terminology is very similar to the one used to deal with the KP hierarchy in the Sato's approach. However, it is worth mentioning that tau functions of the DS hierarchies of $\g$-type in general are \textup{not} KP tau functions (except for $\g=\sl_{n+1}(\CC)$). One way to see it (which is close to the spirit of this paper) is that the generalized Schur polynomials $s_\nu$ of $(\g,\pi)$-type we defined are ``reductions'' (in the sense of the Remark~\ref{1.3}) of the usual ones \cite{Macdonald} just in the $A_n$ case.
\end{Remark}

\begin{Remark}The formula~\eqref{SZ} is {\it intrinsic} when $\pi$ is taken as the adjoint representation of~$\g$. Namely, for such $\pi$ and for each partition, the corresponding Schur polynomials of $(\g,\pi)$-type only depend on the structure constants of~$\g$. We will study the intrinsic Schur polynomials associated to~$\g$ in a~future publication.
\end{Remark}

\begin{Remark}For the $ABCD$ cases, a result similar to Theorem \ref{main} was obtained in \cite{Zhou2} where a different method was used; see also in \cite{BaYZ} for more details for the $A_n$ case.
\end{Remark}

\noindent
{\bf Organization of the paper.} In Section \ref{s2} we review the Drinfeld--Sokolov hierarchies and their tau functions. In Section~\ref{s3} we prove Theorem~\ref{main}. Some explicit examples and applications are given in Section~\ref{s4}. A list of first few Schur polynomials of $(\g,\pi)$-type for $\g$ of low ranks and particular choices of $\pi$ are given in Appendix~\ref{appendixA}.

\section{Review of the Grassmannian approach to the DS hierarchy} \label{s2}
Denote by $\fb$ the Borel subalgebra of $\g$, i.e., $\fb:= \g^{\leq 0}$, and by $\fn$ the nilpotent subalgebra $\fn:=\g^{<0}$. Define a linear operator $\L$ by
\begin{gather*}
\mathcal{L}: = \p_x + \Lambda+ q (x),
\end{gather*}
where $q(x)\in \fb$.
It is proved by V.G.~Drinfeld and V.V.~Sokolov \cite{DS} that there exists a unique smooth function $U(x)\in \g\big(\lambda^{-1}\big)^{<0}\cap \operatorname{Im} \ad_\Lambda$ such that
\begin{gather*}
e^{-\ad_{U(x)}} \L = \p_x+ \Lambda + H(x) ,\qquad H(x)\in \Ker \ad_\Lambda .
\end{gather*}
Here $\g\big(\lambda^{-1}\big)$ consists of formal Laurent series in~$\lambda^{-1}$ with coefficients in~$\g$, and ${}^{<0}$ means taking the subspace of elements with negative principal degrees (the principal degree for elements of~$\g\big(\lambda^{-1}\big)$ is defined similarly as for the loop algebra). The following commuting system of PDEs{\samepage
\begin{gather}\label{pre-flows}
\frac{\p \L}{\p t_\ell} = -\bigl[ \bigl(e^{\ad_U} \Lambda_{\ell}\bigr)_{\geq 0} , \L\bigr] ,\qquad \ell\in E_+
\end{gather}
is called the pre-DS hierarchy of $\g$-type. }

\noindent
{\bf Gauge transformations.} For any smooth function $N(x)\in \fn$, the map
\begin{gather*}
\L \ \mapsto \ \widetilde{\L}= e^{\ad_N} \L = \p_x+ \Lambda+ \tilde q
\end{gather*}
is called a gauge transformation. A vector space $V \subset \g$ is called a {\it DS gauge} if it satisfies
\begin{gather*}
[I_+, \fn] \oplus V = \fb .
\end{gather*}
Below we fix $V$ a DS gauge. It was observed in~\cite{DS} that the flows~\eqref{pre-flows} can be reduced to gauge equivalent classes; moreover, for any $q(x)\in \fb$, there exists a unique~$N(x)$ such that $\tilde q(x) \in V$. Let us denote
\begin{gather*}
\L^{{\rm can}}:= \p_x + \Lambda + q^{{\rm can}}(x) ,\qquad q^{{\rm can}}(x) \in V .
\end{gather*}
Take $v_1,\dots,v_n$ a homogeneous basis of $V$, namely $\deg v_i = -m_i$, and write
\begin{gather*}
q^{{\rm can}} (x)= \sum_{i=1}^n u^i(x) v_i .
\end{gather*}
The DS hierarchy of $\g$-type is defined as the system of the pre-DS flows for the complete set of representatives (aka gauge invariants) $u^1,\dots,u^n$. Clearly, the precise form of this integrable hierarchy depends\footnote{It also depends on scalings of the basis $v_i$ which gives rise to scalings of $u^i$. Such a coordinate change is trivial (in the case $\g=D_{{\rm even}}$ another linear transformation of $u_i$ needs to be considered but is again trivial).} on the choice of the DS gauge~$V$. The hierarchies under different choices of $V$ are Miura equivalent. (To see this we notice that they are all Miura equivalent to the hierarchy under the modified DS gauge, cf.\ Lemma~6.7 of~\cite{DS}, or alternatively they are all Miura equivalent to the DS hierarchy written in the normal coordinates, cf.\ Section~2.8 of~\cite{BDY3}; see also \cite{GHM, HM, HMG}; for the notion of Miura transformation, see~\cite{ DZ-norm}.) We remark that a unified algorithm of writing the DS hierarchy of $\g$-type for an arbitrary choice $V$ was obtained recently in \cite{BDY3}; it has the form
\begin{gather}\label{DSg}
\frac{\p u^i}{\p t_\ell} = a^i_{\ell} \big[u^1,\dots,u^n\big] , \qquad \ell\in E_+,
\end{gather}
where $a_{i,\ell}\big[u^1,\dots,u^n\big]$ are differential polynomials of $u^1,\dots,u^n$. It should also be noted that for the DS hierarchy of $\g$-type the time variable $t_1$ can be identified with~$-x$.

The hierarchy~\eqref{DSg} is known to be tau-symmetric \cite{BDY3, DZ-norm,HM,Wu}. In the setting of~\cite{BDY1,BDY3}, that means that, there exist a family of differential polynomials $\Omega_{k,\ell}$ of $q^{\rm can}$, indexed by two integers $k,\ell\in E_+$ (satisfying certain natural non-degeneracy condition), such that for all $m,k,\ell\in E_+$,
\begin{gather*}
 \Omega_{k,\ell} = \Omega_{\ell,k} , \qquad \frac{\p \Omega_{k,\ell}}{\p t_m} = \frac{\p \Omega_{\ell,m}}{\p t_k} = \frac{\p \Omega_{m,k}}{\p t_\ell} .
\end{gather*}
Therefore, for an arbitrary solution~$q^{{\rm can}}$ of \eqref{DSg}, there exists a function $\tau(\bdt)$ such that
\begin{gather*}
\frac{\p^2 \log \tau}{\p t_k \p t_\ell} = \Omega_{k,\ell} .
\end{gather*}
The function $\tau(\bdt)$ is called the tau-function of the solution~$q^{{\rm can}}$. The tau function is determined by~$q^{{\rm can}}$ up to a multiplicative factor of the form
\begin{gather*}
\exp\biggl( \sum_{\ell\in E_+} c_\ell t_\ell \biggr) ,
\end{gather*}
where $c_\ell$ are arbitrary constants. We review in the rest of this section the Grassmannian approach to tau functions.

Recall that we will fix an irreducible, faithful representation $\pi\colon \mathfrak{g} \to \mathrm{gl}(m,\CC)$ (as in~\eqref{repi}). Let $H := \CC^m \big( \lambda^{-1}\big)$ be the linear space of $\CC^m$-valued formal Laurent series in $\lambda^{-1}$, that is,
\begin{gather*}
H = \lambda^{-1}\CC^m\big[\big[\lambda^{-1}\big]\big] \oplus \CC^m[\lambda] .
\end{gather*}
Let $H_+:=\CC^m[\lambda]$. Denote by ${\rm Gr}$ the Sato--Segal--Wilson Grassmannian~\cite{Sato,SW}, and by $e_1,\dots,e_m$ the canonical basis of~$\CC^m$. A point $W\in{\rm Gr}$ is a subspace of~$H$. Here we are interested in the {\it big cell} ${\rm Gr}^{(0)}\subset{\rm Gr}$ which consists of points $W$ of the form
\begin{gather*}
W= {\rm Span}_{\CC} \biggl\{ e_i \lambda^{\ell} + \sum_{k\geq 0} A_{k,\ell,i} e_i \lambda^{-k-1} \biggr\}_{i=1,\dots,m, \ell\geq0} .
\end{gather*}
Here $A_{k,\ell,i}\in\CC$ are called the affine coordinates \cite{EH} of $W$.

\begin{Definition}Define ${\rm Gr}^{(0)}_{\g}$ as the following subset of the big cell ${\rm Gr}^{(0)}$
\begin{gather*}
{\rm Gr}^{(0)}_{\g} =\Bigl\{ e^a H_+ \,\big| \, a \in \lambda^{-1}\g\big[\lambda^{-1} \big] \Bigr\} .
\end{gather*}
We call ${\rm Gr}^{(0)}_{\g}$ the embedded big cell of $\g$-type.
\end{Definition}

For $a \in \lambda^{-1}\g\big[\lambda^{-1}\big]$, write $G=e^a =\sum\limits_{k\geq 0} G_k \lambda^{-k}$. The matrices $G_0,G_1,\dots$ serve as the matrix-valued coordinates for the point $W$ corresponding to $a$; see Fig.~\ref{MVC}. Clearly, $G_0=I$.

\begin{figure}[htb!]\centering
\begin{tikzpicture}[scale=0.5]
\draw[very thick] (0,0)--(4,0);
\draw (0,2) rectangle (2,4);
\draw (0,0) rectangle (2,2);
\draw (0,-2) rectangle (2,0);
\node at (1,5) {$\vdots$};
\node at (3,5) {$\vdots$};
\node at (5,5) {$\iddots$};
\node at (5,3) {$\cdots$};
\node at (5,1) {$\cdots$};
\node at (5,-1) {$\cdots$};
\node at (5,-3) {$\cdots$};
\node at (5,-4.5) {$\ddots$};
\node at (1,3) {$G_2$};
\node at (1,1) {$G_1$};
\node at (1,-1) {$G_0$};
\draw (2,2) rectangle (4,4);
\draw (2,0) rectangle (4,2);
\draw (2,-2) rectangle (4,0);
\draw (2,-4) rectangle (4,-2);
\node at (3,3) {$G_3$};
\node at (3,1) {$G_2$};
\node at (3,-1) {$G_1$};
\node at (3,-3) {$G_0$};
\end{tikzpicture}
\caption{Matrix-valued coordinates in Sato--Segal--Wilson Grassmannian.}\label{MVC}
\end{figure}

\begin{Definition} Let $M = \sum\limits_{k\in \mathbb{Z}} M_k \lambda^k$ with $M_k\in{\rm gl}(m,\mathbb{C})$. The $N$-th, $N\geq0$, block Toeplitz matrix associated to $M$ is defined by
\begin{gather*}
T_N(M)= (M_{I-J})_{I,J=0}^N .
\end{gather*}
\end{Definition}

 The following theorem comes from the results obtained in \cite{CW1,CW2}.

\begin{Theorem}[Cafasso--Wu \cite{CW1,CW2}] \label{theoremA} For any $X \in \lambda^{-1}\g\big[\lambda^{-1}\big] $, let $\gamma=e^{\xi} e^X$. Define $\tau=\tau(\bdt)$ by
\begin{gather} \label{CWlim}
\tau=\Bigl[\lim_{N\rightarrow \infty } \det T_N(\gamma)\Bigr]^\kappa ,
\end{gather}
where $\kappa$ is defined in \eqref{defkappa}. Then $\tau$ is a tau function of the DS hierarchy associated to~$\g$.
\end{Theorem}

\begin{Remark}
The stabilization proved in \cite{IZ} for the case of the Witten--Kontsevich tau function and extended in \cite{CW2} for the general cases ensures that the limit in \eqref{CWlim} is meaningful.
\end{Remark}

\section{Proof of Theorem~\ref{main}} \label{s3}
\begin{proof}
Define $\gamma= e^{\xi} e^X$, where we recall that $X$ is the given element in $\lambda^{-1}\g\bigl[\bigl[\lambda^{-1}\bigr]\bigr]$, and $\xi = \sum\limits_{ \ell \in E_+} t_\ell \Lambda_\ell$. We have
\begin{gather*}
\mathbb{L}(\gamma)= \mathbb{L}\bigl( e^{\xi} \bigr) \mathbb{L}\big(e^X\big) = s r_X .
\end{gather*}
Here $s$ and $r_X$ are defined in \eqref{defs} and \eqref{defr}, respectively. For any $N\geq 1$, define two matrices
\begin{gather*}s_N = ({s_N}_{, ij})_{i\in\{0,\dots,N\}, j\in\{-N-1,\dots, N\}}\end{gather*} and
\begin{gather*}r_N = ({r_N}_{, ij})_{i\in\{-N-1,\dots, N\}, j\in\{0,\dots,N\}}\end{gather*} by
\begin{gather*}
{s_N}_{, ij} := \mathbb{L}\bigl( e^{\xi} \bigr)_{ij} ,\qquad {r_N}_{, ij} := \mathbb{L}\bigl(e^X\bigr)_{ij} .
\end{gather*}
Then we have
\begin{gather*}
\lim_{N\rightarrow \infty} \det T_N(\gamma) = \lim_{N\rightarrow \infty} \det (s_N r_N ) .
\end{gather*}
Note that $\tau^{1/\kappa}$ is a formal power series in $t_\ell$, $\ell\in E_+$, and the meaning of the above two limits is in the topology of graded formal power series. By using the well-known Cauchy--Binet formula (see for instance~\cite{Gantmacher}) we obtain \cite{EH, Sato} from Theorem~\ref{theoremA} that
\begin{gather*}
\tau^{1/\kappa} = \sum_{\nu \in \mathbb{Y}} r_{X,\nu} s_\nu,
\end{gather*}
where we recall that $r_{X,\nu}$ and $s_\nu$ are defined by
\begin{gather*}r_{X,\nu} = \det (r_{i-\nu_i-1, j-1} )_{i,j=1}^{\ell(\nu)} \qquad \text{and} \qquad s_\nu = \det (s_{i-1, j-\nu_j-1} )_{i,j=1}^{\ell(\nu)} .\end{gather*}
 As explained in \cite{BaY}, formulae \eqref{ms} and \eqref{mac} give the Gaussian eliminations and formulae \eqref{sz1} and \eqref{sz2} are due to the Giambelli-type formula \cite{BaY, EH,Macdonald}. The theorem is proved. \end{proof}

\begin{Remark}Let us mention three important features of Theorem~\ref{main}: i)~For every partition~$\nu$, the function $s_\nu$ is independent of the choice of solutions of the DS hierarchy. ii)~Each summand of~$\sum\limits_{\nu \in \mathbb{Y}} r_{X,\nu} s_\nu$ is accurate; in other words, the limiting procedure is dropped. iii). The expressions~\eqref{sz1} and~\eqref{sz2} give rise to an efficient algorithm for computing tau-functions, as explained in~\cite{BaY,Zhou}.
\end{Remark}

\section{Polynomial tau functions and bilinear equations} \label{s4}
Theorem~\ref{main} gives a simple procedure for efficient computation of tau function $\tau$ when $\tau^{1/\kappa}$ is a~polynomial. Indeed, let us take a faithful representation $\pi$ of $\g$. Choose $X \in \lambda^{-1}\g\big[\big[\lambda^{-1}\big]\big]$ such that $\pi(X)$ is a nilpotent matrix; the infinite series in~\eqref{SZ} becomes finite, as it is easy to verify that only finitely many Pl\"ucker coordinates $\{r_\nu, \, \nu \in \mathbb Y\}$ are non zero. Consequently, $\tau^{1/\kappa}$~is polynomial. This simple idea was used for example in~\cite{BaY} for the KdV hierarchy. If $\kappa = 1$, then the tau function itself is a polynomial. Interestingly enough, in the computations that we will perform, even when $\kappa = 1/2$, we obtain some polynomial tau functions: in other words, the finite sum in~\eqref{SZ} is a perfect square. Even if this result has not been proved in general, we expect that our procedure, under suitable choices of the faithful representation, gives a way of computing all the polynomials tau functions (up to a shift of the times $\{t_i, \, i \in E_+\}$) of the DS hierarchy of $\g$-type. As stated in the introduction of~\cite{KW}, this is an interesting open problem.

\looseness=-1 In what follows we compute the first few polynomial tau functions of the DS hierarchy of $\g$-type for $\g = A_1, A_2, B_2$ and $D_4$. Note that our computations of tau functions (and so of the corresponding solutions to the DS hierarchy) do not require the precise expressions of the equations that we are solving. And note that deriving the explicit expressions of the PDEs in the DS hierarchy (or of their bilinear forms) are themselves very interesting and important questions. These motivate us to do a further application that we explain in more details right below.

We will use the particular tau functions to deduce possible bilinear equations of small degrees. Note that each Drinfeld--Sokolov hierarchy has infinitely many solutions. The usual question is to find particular solutions (and their tau-functions) to the DS hierarchy (e.g., to solve all PDEs in this hierarchy together). Here, as we mentioned above, we will also consider the inverse:
\begin{gather*}
\mbox{Deduce possible bilinear forms of the PDEs from particular solutions}.
\end{gather*}
Sometimes, one particular solution already contains all the information of an equation and of the whole hierarchy. For example, the ``topological solution" was used by B.~Dubrovin and Y.~Zhang to construct the integrable hierarchy of topological type \cite{Du1,Du2, DZ-norm}. However, a polynomial tau function~$\tau_{\rm poly}$ of the DS hierarchy (or say the corresponding solution) contains less information, namely, if $\tau_{\rm poly}$ satisfies some equation, it will not guarantee directly that other tau functions of the DS hierarchy satisfy the same equation. Nevertheless, if $\tau_{\rm poly}$ does not satisfy some equation, then this equation cannot belong to the bilinear forms of the DS hierarchy.

\subsection{Bilinear derivatives}
Given two smooth functions $f(\bdx)$, $g(\bdx)$ with independent variables $\bdx=(x_i)_{i\in I}$, where $I$ denotes an index set. The bilinear derivatives~\cite{Hir} $D_{i_1} \cdots D_{i_k}$ are operators defined via the identity
\begin{gather*}
e^{\sum\limits_{i\in I} h_i D_i}(f,g) \equiv f(\bdx+{\bf h}) g(\bdx-{\bf h}) , \qquad \forall\, {\bf h} .
\end{gather*}
It means that, expanding both sides of this identity in ${\bf h}$
\begin{gather*}
 e^{\sum\limits_{i\in I} h_i D_i}(f,g) = (f,g) + \sum_{i\in I} h_i D_i(f,g) + \sum_{i,j\in I} \frac{h_i h_j}{2}D_iD_j(f,g) + \cdots , \\
 f(\bdx+{\bf h}) g(\bdx-{\bf h}) = f(x) g(x) + \sum_{i\in I} h_i \left( \frac{\partial f}{\partial x_i}g - f\frac{\partial g}{\partial x_i} \right) + \cdots
\end{gather*}
and comparing the coefficients of monomials of ${\bf h}$, we obtain, for example,
\begin{gather*}
 D_i(f,g) = \frac{\partial f}{\partial x_i}g - f\frac{\partial g}{\partial x_i} ,\\
 D_iD_j (f,g) = \frac{\partial^2 f}{\partial x_i \partial x_j}g + f\frac{\partial^2 g}{\partial x_i \partial x_j}
- \frac{\partial f}{\partial x_i}\frac{\partial g}{\partial x_j} - \frac{\partial f}{\partial x_j}\frac{\partial g}{\partial x_i} .
\end{gather*}

For the Drinfeld--Sokolov hierarchy of $\g$-type, we take $I:=E_+$. There is a natural gradation for the bilinear derivatives, defined by assigning $\deg D_i = i$ for $i\in E_+$. Denote by $ \mathcal{H}_{\g} $ the linear space of bilinear equations satisfied by the Drinfeld--Sokolov hierarchies of $\g$-type, which decomposes into homogeneous subspaces
\begin{gather*}
\H_{\g} = \bigoplus_i \H_{\g}^{[i]} .
\end{gather*}
The gradation allows us to list all possible bilinear equations up to a certain degree.

\subsection{Examples of polynomial tau functions}
\subsubsection[The $A_1$ case]{The $\boldsymbol{A_1}$ case}
Let us chose the standard matrix realization $\g=\sl(2;\mathbb{C})$. Consider the following two elements in $ \lambda^{-1}\g\bigl[\bigl[\lambda^{-1}\bigr]\bigr]$
\begin{gather*}
\frac{1}\lambda F = \frac{1}{\lambda}\begin{pmatrix}
0&0\\
1&0
\end{pmatrix}, \qquad \frac{1}{\lambda} E = \frac{1}{\lambda}\begin{pmatrix}
0&1\\
0&0
\end{pmatrix}.
\end{gather*}
The associated polynomial tau functions are
\begin{gather}
\tau_1 = 1 + t_1 , \qquad \tau_2 = 1 +t_3 - \frac{t_1^3}{3} , \label{tau1-2KdV}
\end{gather}
respectively. Similarly, one computes polynomial tau functions corresponding to elements of the form $\lambda^{-k} F $, $\lambda^{-k} E$, $k\geq 2$. For example, for $k=2$, we obtain
\begin{gather}
 \tau_3 = 1+2 t_3-t_5 t_1+t_3^2+\frac{t_1^3}{3}+\frac{1}{3} t_3 t_1^3-\frac{1}{45} t_1^6 , \label{tau3KdV}\\
 \tau_4 = 1-t_3t_7+2 t_5+t_5^2+t_3^3 t_1-t_3 t_5 t_1^2-t_3 t_1^2+\frac{1}{3} t_7 t_1^3-\frac{t_1^5}{15}\nonumber\\
 \hphantom{\tau_4 =}{} -\frac{1}{15} t_5 t_1^5+\frac{1}{105} t_3 t_1^7-\frac{t_1^{10}}{4725} , \label{tau4KdV}
\end{gather}
corresponding to $\lambda^{-2} F$ and $\lambda^{-2} E$, respectively.

Now consider all bilinear equations up to degree 4
\begin{gather}
\big(\beta + \alpha_0 D_1^2 + \alpha_1 D_1^4 + \alpha_2 D_1D_3\big)(\tau,\tau) = 0 , \label{ana1-1}
\end{gather}
where $\beta$, $\alpha_0$, $\alpha_1$, $\alpha_2$ are complex constants. Requiring that $\tau_1$, $\tau_2$ satisfy the above ansatz~\eqref{ana1-1}, we find that up to a multiplicative constant there is only one possible choice of coefficients:
\begin{gather} \big(D_1^4 - 4D_1D_3\big)(\tau,\tau) = 0 . \label{kdv1}\end{gather}
Similarly up to degree 6, we find out only two more possible linearly independent bilinear equations that are satisfied by $\tau_1$, $\tau_2$, $\tau_3$, $\tau_4$
\begin{gather}
 \big(D_1^6 + 20D_1^3D_3 - 96D_1D_5\big)(\tau,\tau) = 0 , \label{kdv21} \\
 \big(D_1^3D_3 + 2 D_3^2 - 6 D_1D_5\big)(\tau,\tau) =0 ,\label{kdv22}
\end{gather}
which are identified to two of the well-known bilinear equations for the hierarchy of $A_1$-type (the KdV hierarchy). Consequently, we have shown that
\begin{gather*}\dim_{\CC} \mathcal{H}^{[\deg \leq 6]}_{A_1} \leq 3 .\end{gather*}
Moreover, \eqref{kdv1}--\eqref{kdv22} are the three only possible choices of homogeneous basis (up to constant factors) of $\mathcal{H}^{[\deg \leq 6]}_\g$.

\medskip

\noindent
\textbf{Relation with the Adler--Moser polynomials.} An alternative way of computing polynomial tau functions for the KdV hierarchy was given by Adler and Moser~\cite{AdMPoly}. Define a family of polynomials $\theta_k(x=q_1,q_3,q_5,\ldots,q_{2k-1})$, $k\geq 0$, recursively by
\begin{gather*}
 \theta_0 = 1 , \qquad \theta_1 = x , \qquad \theta_{k+1}'\theta_{k-1} + \theta_{k+1}\theta_{k-1}' = (2k-1)\theta_k^2 , \qquad \forall\, k \geq 2 ,
\end{gather*}
where the prime denotes the $x$-derivative and for each $k\geq 2$ the integration constant is chosen to be $q_{2k-1}$. The polynomials $\theta_k$ are known as the Adler--Moser polynomials. It was also proven in \cite{AdMPoly} that there exists a unique change of variables $\mathbf{q}\to \mathbf{t}$ that transforms the Adler--Moser polynomials into the polynomial tau functions of the KdV hierarchy. In~\cite{Duc}, one of the authors of the present paper proved that the desired change of variables is given by $q_1 = t_1 = x$ and
\begin{gather*}
\sum_{i\geq 2} \frac{q_{2i-1}}{\alpha_{2i-1}}z^{2i-1} = \tanh\biggl( \sum_{i\geq 2} t_{2i-1} z^{2i-1} \biggr) ,
\end{gather*}
where $\alpha_{2i-1} := (-1)^{i-1}3^2\cdots (2i-3)^2(2i -1)$. Up to a shift and renormalisation of the times, we recover in particular the polynomials given in equations \eqref{tau1-2KdV}--\eqref{tau4KdV}.

\subsubsection[The $A_2$ case]{The $\boldsymbol{A_2}$ case}
We still chose the standard matrix realization $\g= \sl(3;\mathbb{C})$. Consider for example the following two elements in $ \lambda^{-1}\g\bigl[\bigl[\lambda^{-1}\bigr]\bigr]$:
\begin{gather*} X_1= \frac{1}{\lambda} \begin{pmatrix}
0 & 0 & 0\\
a_1 & 0 &0\\
a_2 & a_3& 0
\end{pmatrix}, \qquad X_2= \frac{1}{\lambda} \begin{pmatrix}
0& a_1& a_2\\
0 & 0 & a_3\\
0&0&0
\end{pmatrix}, 
\end{gather*}
where $a_1$, $a_2$, $a_3$ are arbitrary constants. The corresponding polynomial tau functions will be denoted by $\tau_1$, $\tau_2$, respectively. We have
\begin{gather*}
\tau_1 = 1+{a_2} t_1+\frac{1}{2} {a_1} t_1^2-\frac{1}{2} {a_3} t_1^2+\frac{1}{8} {a_1} {a_3} t_1^4-\frac{1}{160} a_1^2 {a_3} t_1^6+\frac{1}{160} {a_1} a_3^2 t_1^6-\frac{ a_1^2 a_3^2 t_1^8}{1792}+{a_1} t_2\\
\hphantom{\tau_1 =}{} +{a_3} t_2+\frac{1}{16} a_1^2 {a_3} t_1^4 t_2+\frac{1}{16} {a_1} a_3^2 t_1^4 t_2+\frac{3}{2} {a_1} {a_3} t_2^2-\frac{1}{8} a_1^2 {a_3} t_1^2 t_2^2+\frac{1}{8} {a_1} a_3^2 t_1^2 t_2^2+\frac{1}{32} a_1^2 a_3^2 t_1^4 t_2^2\\
\hphantom{\tau_1 =}{} +\frac{1}{4} a_1^2 {a_3} t_2^3+\frac{1}{4} {a_1} a_3^2 t_2^3+\frac{1}{16} a_1^2 a_3^2 t_2^4-\frac{1}{4} a_1^2 {a_3} t_1^2 t_4-\frac{1}{4} {a_1} a_3^2 t_1^2 t_4-\frac{1}{2} a_1^2 {a_3} t_2 t_4+\frac{1}{2} {a_1} a_3^2 t_2 t_4\\
\hphantom{\tau_1 =}{} -\frac{1}{4} a_1^2 a_3^2 t_1^2 t_2 t_4-\frac{1}{4} a_1^2 a_3^2 t_4^2+\frac{1}{2} a_1^2 {a_3} t_1 t_5-\frac{1}{2} {a_1} a_3^2 t_1 t_5+\frac{1}{4} a_1^2 a_3^2 t_1 t_7 ,
\\
\tau_2 = 1-\frac{1}{8} {a_1} t_{1}^4+\frac{1}{8} {a_3} t_{1}^4+\frac{1}{20} {a_2} t_{1}^5+\frac{1}{640} {a_1} {a_3} t_{1}^8-\frac{ a_1^2 {a_3} t_{1}^{12}}{358400}+\frac{{a_1} a_3^2 t_{1}^{12}}{358400}-\frac{ a_1^2 a_3^2 t_{1}^{16}}{90112000}-\frac{1}{2} {a_1} t_{1}^2 t_{2}\\
\hphantom{\tau_2 =}{} -\frac{1}{2} {a_3} t_{1}^2 t_{2}-\frac{ a_1^2 {a_3} t_{1}^{10} t_{2}}{12800}-\frac{{a_1} a_3^2 t_{1}^{10} t_{2}}{12800}+\frac{1}{2} {a_1} t_{2}^2-\frac{1}{2} {a_3} t_{2}^2-{a_2} t_{1} t_{2}^2+\frac{1}{16} {a_1} {a_3} t_{1}^4 t_{2}^2\\
\hphantom{\tau_2 =}{} -\frac{13 a_1^2 {a_3} t_{1}^8 t_{2}^2}{17920}+\frac{13 {a_1} a_3^2 t_{1}^8 t_{2}^2}{17920}+\frac{3 a_1^2 a_3^2 t_{1}^{12} t_{2}^2}{1126400}-\frac{1}{320} a_1^2 {a_3} t_{1}^6 t_{2}^3-\frac{1}{320} {a_1} a_3^2 t_{1}^6 t_{2}^3-\frac{3}{8} {a_1} {a_3} t_{2}^4\\
\hphantom{\tau_2 =}{}-\frac{1}{128} a_1^2 {a_3} t_{1}^4 t_{2}^4+\frac{1}{128} {a_1} a_3^2 t_{1}^4 t_{2}^4-\frac{ a_1^2 a_3^2 t_{1}^8 t_{2}^4}{10240}-\frac{1}{32} a_1^2 {a_3} t_{1}^2 t_{2}^5-\frac{1}{32} {a_1} a_3^2 t_{1}^2 t_{2}^5-\frac{1}{32} a_1^2 {a_3} t_{2}^6\\
\hphantom{\tau_2 =}{} +\frac{1}{32} {a_1} a_3^2 t_{2}^6+\frac{1}{256} a_1^2 a_3^2 t_{1}^4 t_{2}^6+\frac{1}{256} a_1^2 a_3^2 t_{2}^8+{a_1} t_{4}+{a_3} t_{4}+\frac{ a_1^2 {a_3} t_{1}^8 t_{4}}{1280}+\frac{{a_1} a_3^2 t_{1}^8 t_{4}}{1280}\\
\hphantom{\tau_2 =}{} -\frac{3}{2} {a_1} {a_3} t_{1}^2 t_{2} t_{4}+\frac{1}{160} a_1^2 {a_3} t_{1}^6 t_{2} t_{4}-\frac{1}{160} {a_1} a_3^2 t_{1}^6 t_{2} t_{4}-\frac{ a_1^2 a_3^2 t_{1}^{10} t_{2} t_{4}}{12800}+\frac{1}{32} a_1^2 {a_3} t_{1}^4 t_{2}^2 t_{4}\\
\hphantom{\tau_2 =}{}+\frac{1}{32} {a_1} a_3^2 t_{1}^4 t_{2}^2 t_{4}-\frac{1}{8} a_1^2 {a_3} t_{1}^2 t_{2}^3 t_{4}+\frac{1}{8} {a_1} a_3^2 t_{1}^2 t_{2}^3 t_{4}-\frac{1}{320} a_1^2 a_3^2 t_{1}^6 t_{2}^3 t_{4}-\frac{3}{16} a_1^2 {a_3} t_{2}^4 t_{4}\\
\hphantom{\tau_2 =}{} -\frac{3}{16} {a_1} a_3^2 t_{2}^4 t_{4}-\frac{1}{32} a_1^2 a_3^2 t_{1}^2 t_{2}^5 t_{4}+\frac{3}{2} {a_1} {a_3} t_{4}^2+\frac{1}{32} a_1^2 {a_3} t_{1}^4 t_{4}^2-\frac{1}{32} {a_1} a_3^2 t_{1}^4 t_{4}^2+\frac{ a_1^2 a_3^2 t_{1}^8 t_{4}^2}{2560}\\
\hphantom{\tau_2 =}{} -\frac{3}{8} a_1^2 {a_3} t_{1}^2 t_{2} t_{4}^2-\frac{3}{8} {a_1} a_3^2 t_{1}^2 t_{2} t_{4}^2-\frac{1}{8} a_1^2 {a_3} t_{2}^2 t_{4}^2+\frac{1}{8} {a_1} a_3^2 t_{2}^2 t_{4}^2+\frac{1}{64} a_1^2 a_3^2 t_{1}^4 t_{2}^2 t_{4}^2-\frac{3}{32} a_1^2 a_3^2 t_{2}^4 t_{4}^2\\
\hphantom{\tau_2 =}{} +\frac{1}{4} a_1^2 {a_3} t_{4}^3+\frac{1}{4} {a_1} a_3^2 t_{4}^3-\frac{1}{8} a_1^2 a_3^2 t_{1}^2 t_{2} t_{4}^3+\frac{1}{16} a_1^2 a_3^2 t_{4}^4+{a_2} t_{5}+\frac{1}{2} {a_1} {a_3} t_{1}^3 t_{5}+\frac{3 a_1^2 {a_3} t_{1}^7 t_{5}}{1120}\\
\hphantom{\tau_2 =}{} -\frac{3 {a_1} a_3^2 t_{1}^7 t_{5}}{1120}+\frac{ a_1^2 a_3^2 t_{1}^{11} t_{5}}{140800}+\frac{1}{80} a_1^2 {a_3} t_{1}^5 t_{2} t_{5}+\frac{1}{80} {a_1} a_3^2 t_{1}^5 t_{2} t_{5}+\frac{1}{8} a_1^2 {a_3} t_{1}^3 t_{2}^2 t_{5}-\frac{1}{8} {a_1} a_3^2 t_{1}^3 t_{2}^2 t_{5}\\
\hphantom{\tau_2 =}{} +\frac{1}{320} a_1^2 a_3^2 t_{1}^7 t_{2}^2 t_{5}+\frac{1}{4} a_1^2 {a_3} t_{1} t_{2}^3 t_{5}+\frac{1}{4} {a_1} a_3^2 t_{1} t_{2}^3 t_{5}-\frac{1}{32} a_1^2 a_3^2 t_{1}^3 t_{2}^4 t_{5}+\frac{1}{4} a_1^2 {a_3} t_{1}^3 t_{4} t_{5}\\
\hphantom{\tau_2 =}{} +\frac{1}{4} {a_1} a_3^2 t_{1}^3 t_{4} t_{5}-\frac{1}{2} a_1^2 {a_3} t_{1} t_{2} t_{4} t_{5}+\frac{1}{2} {a_1} a_3^2 t_{1} t_{2} t_{4} t_{5}+\frac{1}{80} a_1^2 a_3^2 t_{1}^5 t_{2} t_{4} t_{5}+\frac{1}{4} a_1^2 a_3^2 t_{1} t_{2}^3 t_{4} t_{5}\\
\hphantom{\tau_2 =}{} +\frac{1}{8} a_1^2 a_3^2 t_{1}^3 t_{4}^2 t_{5}+\frac{1}{4} a_1^2 {a_3} t_{1}^2 t_{5}^2-\frac{1}{4} {a_1} a_3^2 t_{1}^2 t_{5}^2-\frac{1}{160} a_1^2 a_3^2 t_{1}^6 t_{5}^2-\frac{1}{2} a_1^2 {a_3} t_{2} t_{5}^2-\frac{1}{2} {a_1} a_3^2 t_{2} t_{5}^2\\
\hphantom{\tau_2 =}{} -\frac{1}{8} a_1^2 a_3^2 t_{1}^2 t_{2}^2 t_{5}^2-\frac{1}{2} a_1^2 a_3^2 t_{2} t_{4} t_{5}^2-\frac{1}{4} a_1^2 a_3^2 t_{1} t_{5}^3-\frac{1}{40} a_1^2 {a_3} t_{1}^5 t_{7}+\frac{1}{40} {a_1} a_3^2 t_{1}^5 t_{7}+\frac{1}{2} a_1^2 {a_3} t_{1} t_{2}^2 t_{7}\\
\hphantom{\tau_2 =}{} -\frac{1}{2} {a_1} a_3^2 t_{1} t_{2}^2 t_{7}-\frac{1}{2} a_1^2 {a_3} t_{5} t_{7}+\frac{1}{2} {a_1} a_3^2 t_{5} t_{7}-\frac{1}{16} a_1^2 {a_3} t_{1}^4 t_{8}-\frac{1}{16} {a_1} a_3^2 t_{1}^4 t_{8}-\frac{1}{4} a_1^2 {a_3} t_{1}^2 t_{2} t_{8}\\
\hphantom{\tau_2 =}{} +\frac{1}{4} {a_1} a_3^2 t_{1}^2 t_{2} t_{8}-\frac{1}{160} a_1^2 a_3^2 t_{1}^6 t_{2} t_{8}+\frac{1}{4} a_1^2 {a_3} t_{2}^2 t_{8}+\frac{1}{4} {a_1} a_3^2 t_{2}^2 t_{8}+\frac{1}{8} a_1^2 a_3^2 t_{1}^2 t_{2}^3 t_{8}+\frac{1}{2} a_1^2 {a_3} t_{4} t_{8}\\
\hphantom{\tau_2 =}{} -\frac{1}{2} {a_1} a_3^2 t_{4} t_{8}-\frac{1}{16} a_1^2 a_3^2 t_{1}^4 t_{4} t_{8}+\frac{1}{4} a_1^2 a_3^2 t_{2}^2 t_{4} t_{8}+\frac{1}{2} a_1^2 a_3^2 t_{1} t_{2} t_{5} t_{8}-\frac{1}{4} a_1^2 a_3^2 t_{8}^2+\frac{1}{80} a_1^2 a_3^2 t_{1}^5 t_{11}\\
\hphantom{\tau_2 =}{} -\frac{1}{4} a_1^2 a_3^2 t_{1} t_{2}^2 t_{11}+\frac{1}{4} a_1^2 a_3^2 t_{5} t_{11} .
\end{gather*}

Consider all possible bilinear equations of degree 4:
\begin{gather*}\big(\alpha_1D_1^4 + \alpha_2D_2^2\big)(\tau,\tau) = 0.\end{gather*}
Requiring that $\tau_1$ satisfies this ansatz we find that there is only one possible choice:
\begin{gather*}\big(D_1^4 + 3D_2^2\big)(\tau,\tau) = 0 .\end{gather*}
Similarly, requiring that $\tau_1$ and $\tau_2$ both satisfy the ansatz of bilinear equation of degree 6, we find that there are only two linearly independent bilinear equations of degree 6:
\begin{gather*}
\big(D_1^6 + 45D_1^2D_2^2 + 90D_2D_4 - 216D_1D_5\big)(\tau,\tau) = 0 ,\\
\big(D_1^6 + 15D_1^2D_2^2 + 60D_2D_4 - 96D_1D_5\big)(\tau,\tau) = 0 ,
\end{gather*}
which are identified to two of the well-known bilinear equations for the hierarchy of $A_2$-type (the Boussinesq hierarchy).

\subsubsection[The $B_2$ case]{The $\boldsymbol{B_2}$ case}\label{section4.2.3}
We choose the matrix realization of the $B_2$ simple Lie algebra as in~\cite{DS} (cf.\ p.~2032 therein). We consider two explicit examples given respectively by the following matrices\footnote{$X_2$ is not the most general upper triangular element of homogeneous degree~$-1$, as the tau function for the most general case is too big.}
\begin{align*}X_1= \frac{1}{\lambda} \begin{pmatrix}
0 & 0 & 0 & 0 & 0\\
a_2 & 0 & 0 & 0 & 0\\
a_3 & a_5 & 0 & 0 & 0\\
a_4 & 0 & a_5 & 0 & 0\\
0 & a_4 & -a_3 & a_2 & 0
\end{pmatrix} , \qquad X_2= \frac{1}{\lambda} \begin{pmatrix}
0 & 0 & a_3 & a_4 & 0\\
0 & 0 & 0 & 0 & a_4\\
0 & 0 & 0 & 0 & -a_3\\
0 & 0 & 0 & 0 & 0\\
0 & 0 & 0 & 0 & 0
\end{pmatrix}.\end{align*}

The associated tau functions will be denoted by $\tau_1$ and $\tau_2$. They have the expressions
\begin{gather*}
\tau_1 =  1+\frac{a_4}{2} t_{1}+\frac{a_3}{4} t_{1}^2 + \left(\frac{a_2}{12} -\frac{a_5}{12} \right) t_{1}^3
+ \left( \frac{a_2 a_4}{96} -\frac{a_3^2}{192}  \right)t_{1}^4
+\frac{a_3 a_5}{192} t_{1}^5+ \left(\frac{a_2 a_5}{1920}-\frac{a_5^2}{720} \right)  t_{1}^6  \\
\hphantom{\tau_1 =}{} -\frac{a_2 a_4 a_5}{11520} t_{1}^7-\frac{a_2 a_3 a_5}{53760}  t_{1}^8 +
\left(\frac{a_2 a_5^2}{2903040} - \frac{a_2^2 a_5}{483840}\right) t_1^9  \\
\hphantom{\tau_1 =}{} +
(\mbox{more than 300 terms})  \\
\hphantom{\tau_1 =}{} + \left(\frac{a_2^3 a_5^6}{13824} - \frac{a_2^4 a_5^5}{18432}\right)t_3 t_5^2 t_7^2,
\\
\tau_2 = 1+\frac{1}{288} {a_3} t_1^6-\frac{{a_4} t_1^7}{2016}-\frac{ a_3^2 t_1^{12}}{11612160}-\frac{1}{12} {a_3} t_1^3 t_3+\frac{1}{48} {a_4} t_1^4 t_3-\frac{ a_3^2 t_1^9t_3}{69120}+\frac{1}{2} {a_3} t_3^2-\frac{1}{2} {a_4} t_1 t_3^2\\
\hphantom{\tau_2 =}{}-\frac{ a_3^2 t_1^6 t_3^2}{1920}-\frac{1}{48} a_3^2 t_1^3 t_3^3+\frac{1}{16} a_3^2 t_3^4+\frac{ a_3^2 t_1^7 t_5}{4032}-\frac{1}{96} a_3^2 t_1^4 t_3 t_5+\frac{1}{4} a_3^2 t_1 t_3^2 t_5+{a_4} t_7+\frac{1}{160} a_3^2 t_1^5 t_7\\
\hphantom{\tau_2 =}{}-\frac{1}{2} a_3^2 t_5 t_7 .
\end{gather*}

Consider all bilinear equations up to degree 4
\begin{align*}\big(\alpha_0 + \alpha_1 D_1^2 + \alpha_2 D_1^4 + \alpha_3 D_1D_3\big)(\tau,\tau) = 0 , \end{align*}
where $\alpha_0,\dots,\alpha_3$ are constants. Requiring that $\tau_1$ satisfies this ansatz of bilinear equations we find that there is no solution. Similarly, up to degree 8, we find that there are only two possible homogeneous equations (one is of degree $6$ and the other is of degree $8$). We arrive at
\begin{Proposition} The following dimension estimates hold true
\begin{gather*}
\dim_{\CC} \mathcal{H}^{[\deg \leq 4]}_{B_2} = 0 , \qquad \dim_{\CC} \mathcal{H}^{[\deg \leq 6]}_{B_2} \leq 1 , \qquad \dim_{\CC} \mathcal{H}^{[\deg \leq 8]}_{B_2} \leq 2 .
\end{gather*}
Moreover, the only possible elements in $\mathcal{H}^{[\deg \leq 8]}_{B_2} $ are linear combinations of
\begin{gather*}\big(D_1^6 - 5 D_1^3 D_3 - 5 D_3^2 + 9 D_1 D_5\big)(\tau_,\tau) = 0 ,\end{gather*}
and
\begin{gather*}\big(D_1^8 + 7 D_1^5 D_3 - 35 D_1^2 D_3^2 - 21 D_1^3 D_5 - 42 D_3 D_5 + 90 D_1 D_7\big)(\tau,\tau) = 0 .\end{gather*}
\end{Proposition}
\begin{Remark}As far as we know, explicit bilinear equations for the DS hierarchy of $B_2$-type are not pointed out in the literature, except that there is a super-variable version given in~\cite{KW}. However, the relationship between the super bilinear equations of Kac--Wakimoto~\cite{KW} and the DS hierarchy of $B_2$-type is not known. Finding explicit generating series of bilinear equations for the DS hierarchy of $B_2$-type remains an open question. It is also interesting to remark that the very same equations are contained in~\cite{DJKM2}, as the first two equations of the BKP hierarchy~\cite{DJKM2,NO,S,You}.
\end{Remark}

\subsubsection[The $D_4$ case]{The $\boldsymbol{D_4}$ case}
Take the matrix realization of~$\g$ as in~\cite{BDY2} (cf.\ Example~4.4 therein). Consider the particular point of the Sato Grassmannian of $D_4$-type given by
\begin{gather*}
\gamma=1 + \lambda E_{\theta} .
\end{gather*}
We put $t_{11}=0$. It follows from Theorem \ref{main} that the corresponding tau function is given by
\begin{gather*}
\tau = \left(1 - \frac1 2 s_{(7|6)} - \frac12 s_{(6|7)} - \frac 14 s_{(7,6|7,6)}\right)^{\frac12},
\end{gather*}
where $s_{(7,6|7,6)} = s_{(7|7)} s_{(6|6)} - s_{(7|6)} s_{(6|7)} $, $s_{(6|6)} = s_{(7|7)} = 0$, and
\begin{gather*}
 s_{(6|7)} = s_{(7|6)} = \frac{t_1^{11}}{1900800}-\frac{1}{480} t_{5} t_1^6+\frac{1}{160} t_3^2 t_1^5+\frac{1}{120} t_{3'}^2 t_1^5+\frac{1}{80} t_3 t_{3'} t_1^5 \\
\hphantom{s_{(6|7)} = s_{(7|6)} =}{} -\frac{1}{8} t_3^3 t_1^2-\frac{1}{4} t_3 t_{3'}^2 t_1^2-\frac{3}{8} t_3^2 t_{3'} t_1^2+\frac{1}{2} t_{5}^2 t_1 +\frac{3}{4} t_3^2 t_{5}+t_{3'}^2 t_{5}+\frac{3}{2} t_3 t_{3'} t_{5} .
\end{gather*}

Hence we have
\begin{gather*}
\tau = 1- \frac12 s_{(7|6)} = 1 -\frac{t_1^{11}}{3801600}+\frac{1}{960} t_5 t_1^6-\frac{1}{320} t_3^2 t_1^5-\frac{1}{240} t_{3'}^2 t_1^5-\frac{1}{160} t_3 t_{3'} t_1^5 \\
 \hphantom{\tau = 1- \frac12 s_{(7|6)} =}{} +\frac{1}{16} t_3^3 t_1^2+\frac{1}{8} t_3 t_{3'}^2 t_1^2 +\frac{3}{16} t_3^2 t_{3'} t_1^2-\frac{1}{4} t_5^2 t_1-\frac{3}{8} t_3^2 t_5-\frac{1}{2} t_{3'}^2 t_5-\frac{3}{4} t_3 t_{3'} t_5 .
\end{gather*}

\begin{Proposition}The following dimension estimates hold true
\begin{gather*}
\dim_{\CC} \mathcal{H}^{[\deg \leq 4]}_{D_4} = 0 , \qquad \dim_{\CC} \mathcal{H}^{[\deg \leq 6]}_{D_4} \leq 3 .
\end{gather*}
Moreover, the only possible elements in $\mathcal{H}^{[\deg \leq 6]}_{D_4} $ are linear combinations of
\begin{gather}
 \big(2 D_1^3 D_{3'} + 4 D_3 D_{3'} - 3 D_{3'}^2 \big)(\tau,\tau) = 0 ,\label{D4bi1}\\
 \big(D_1^3 D_3 - D_1^3 D_{3'} + D_3 D_{3'} - D_3^2\big)(\tau,\tau) = 0 ,\label{D4bi2}\\
 \big(D_1^6 + 9 D_1 D_5 - 10 D_1^3 D_3 + 5 D_1^3 D_{3'} - 5 D_3 D_{3'} \big)(\tau,\tau) = 0 . \label{D4bi3}
\end{gather}
\end{Proposition}

Our last remark is that under the following linear change of time variables
\begin{gather*}
 \p_{t_1} \mapsto 2^{-1/6} \p_{T_1} , \qquad \p_{t_3} \mapsto 2^{1/2} \p_{T_3} , \qquad \p_{t_{3'}} \mapsto 2^{1/2} \p_{T_3} + 2^{1/2} 3^{-1/2} \p_{T_{3'}} , \qquad
\p_{t_5} \mapsto 2^{7/6} \p_{T_5} ,
\end{gather*}
the bilinear equations \eqref{D4bi1}--\eqref{D4bi3} in the new time variables $T_1$, $T_3$, $T_{3'}$, $T_5$ coincide with those of Kac and Wakimoto~\cite{KW}. Essentially speaking such a change of times is simply a renormalization of flows.

\appendix
\section[List of generalized Schur polynomials of $(\mathfrak{g},\pi)$-type]{List of generalized Schur polynomials of $\boldsymbol{(\mathfrak{g},\pi)}$-type}\label{appendixA}

Take $\pi$ as in~\cite{BDY2, DS} (cf.~\cite[Example~4.4]{BDY2} and \cite[p.~2032]{DS}). We list in Table~\ref{nums} the first several Schur polynomials of $(\g,\pi)$-type for simple Lie algebras of low ranks.
\begin{table}[th]\centering
\begin{tabular} {|M{0.7cm}|M{1.7cm}|M{2.2cm}|M{1.8cm}|M{1.8cm}|M{1.8cm}|M{1.8cm}|N}
 \hline
 $\g$ & $A_1$ & $A_2$ & $B_2$ &$B_3$ & $C_2$ & $D_4$ & \\[10pt]
 \hline
 $s_1$ & $t_1$ & $t_1$	& $0$ & $0$ & $t_1$ 	 & $0$ & \\[8pt]
 \hline
 $s_2$ & $\frac12 {t_1^2}$ & $\frac12 {t_1^2}+t_2$& $\frac12 {t_1}$ &$\frac12 {t_1}$ & $\frac12 t_1^2$ & $\frac12 t_1$ & \\[8pt]
 \hline
 $s_{1^2}$ & $\frac12 {t_1^2}$ & $\frac12 {t_1^2}-t_2$ & $-\frac12 {t_1}$ &$-\frac12 {t_1}$ & $ \frac12 t_1^2$ & $-\frac12 t_1$ & \\[8pt]
 \hline
 $s_3$ & $\frac16 {t_1^3} +t_3$ & $\frac16 {t_1^3}+t_1t_2$ & $\frac14 {t_1^2}$ & $\frac14 {t_1^2}$ & $\frac13 {t_1^3}+2t_3$ & $ \frac14 t_1^2$ & \\[8pt]
 \hline
 $s_{21}$ & $\frac13 {t_1^3} -t_3$ & $\frac13 {t_1^3}$ & $0$ & $0$ & $\frac13 {t_1^3}-t_3$ & $0$ & \\[8pt]
 \hline
 $s_{1^3}$ & $\frac16 {t_1^3} +t_3$ & $\frac16 {t_1^3}-t_1t_2$ & $-\frac14 t_1^2$ & $-\frac14 {t_1^2}$ & $\frac13 {t_1^3}+2t_3$ & $-\frac14 t_1^2$ & \\[8pt]
 \hline
 $s_4$ & $\frac1{24} t_1^4+t_3 t_1$ & $\frac1{24} t_1^4 + \frac{1}{2} t_2 t_1^2 + \frac 12 t_2^2+t_4$
 & $\frac1{12} {t_1^3} +\frac12 t_3$ & $\frac1{12} {t_1^3}+\frac12 {t_3}$ & $\frac1{12} t_1^4+ 2 t_1 t_3$ & $\frac{1}{12}t_1^3+\frac{1}{2}t_3+t_{3'}$ &\\[16pt]
 \hline
 $s_{31}$ & $\frac1 8 t_1^4$ & $\frac18 t_1^4 + \frac12 t_1^2 t_2 - \frac12 t_2^2 - t_4$ & $\frac1{12} t_1^3-t_3$ & $\frac1{12} {t_1^3} - t_3 $ & $\frac14 {t_1^4}$ & $\frac{1}{12}t_1^3-t_3-t_{3'}$ & \\[16pt]
 \hline
 $s_{2^2}$ & $\frac1{12} t_1^4 - t_1 t_3$ & $\frac1{12} t_1^4+t_2^2$ & $\frac14 t_1^2$ & $\frac14 {t_1^2}$ & $\frac1{12} t_1^4 - t_1 t_3$ & $\frac14 t_1^2$ & \\[8pt]
 \hline
 $s_{21^2}$ & $\frac18 t_1^4$ & $\frac18 t_1^4 - \frac{1}{2} t_1^2 t_2-\frac12 t_2^2+t_4$ & $-\frac1{12} {t_1^3} + t_3 $ & $-\frac1{12} {t_1^3} + t_3 $ & $\frac14 {t_1^4}$ & $-\frac{1}{12}t_1^3+t_3+t_{3'}$ &\\[16pt]
 \hline
 $s_{1^4}$ & $\frac1{24} t_1^4+t_3 t_1$ & $\frac1{24} t_1^4 - \frac{1}{2} t_2 t_1^2 + \frac 12 t_2^2 - t_4$
 & $- \frac1{12} {t_1^3} - \frac12 {t_3}$ & $- \frac1{12} {t_1^3}- \frac12 {t_3}$ & $\frac1{12} t_1^4+ 2 t_1 t_3$ & $-\frac{1}{12}t_1^3-\frac{1}{2}t_3-t_{3'}$ & \\[16pt]
 \hline
 \end{tabular}
\caption{Simple Lie algebras and Schur polynomials of $(\g,\pi)$-type.} \label{nums}
\end{table}

\subsection*{Acknowledgements}
We would like to thank Ferenc Balogh, Marco Bertola, Boris Dubrovin, John Harnad, Leonardo Patimo, Daniele Valeri, Chao-Zhong Wu and Jian Zhou for helpful discussions. D.Y. is grateful to Youjin Zhang and Boris Dubrovin for their advisings, and to Victor Kac for helpful suggestions. We thank the anonymous referees for constructive comments.
We would like to thank Fabrizio Del Monte for observing a computational error for $\tau_1$ in Section~\ref{section4.2.3} and
Jindong Guo for his assistance on the new computation.
Part of our work was done at SISSA; we acknowledge SISSA for excellent working conditions and generous supports. A.D.~and M.C.~thank the Centre Henri Lebesgue ANR-11-LABX-0020-01 for creating an attractive mathematical environment. Part of the work of D.Y.~was done during his visits to LAREMA; he acknowledges the support of LAREMA and warm hospitality. A.D.~and M.C.~acknowledge the support of the project IPaDEGAN (H2020-MSCA-RISE-2017), Grant No.~778010.

\pdfbookmark[1]{References}{ref}
\LastPageEnding


\begin{thebibliography}{99}
\footnotesize\itemsep=0pt

\bibitem{AdMPoly}
Adler M., Moser J., On a class of polynomials connected with the {K}orteweg--de
 {V}ries equation, \href{https://doi.org/10.1007/BF01609465}{\textit{Comm. Math. Phys.}} \textbf{61} (1978), 1--30.

\bibitem{BFRFW}
Balog J., Feh\'{e}r L., O'Raifeartaigh L., Forg\'{a}cs P., Wipf A., Toda theory
 and {$\mathcal W$}-algebra from a gauged {WZNW} point of view, \href{https://doi.org/10.1016/0003-4916(90)90029-N}{\textit{Ann.
 Physics}} \textbf{203} (1990), 76--136.

\bibitem{BaY}
Balogh F., Yang D., Geometric interpretation of {Z}hou's explicit formula for
 the {W}itten--{K}ontsevich tau function, \href{https://doi.org/10.1007/s11005-017-0965-8}{\textit{Lett. Math. Phys.}}
 \textbf{107} (2017), 1837--1857, \href{http://arxiv.org/abs/1412.4419}{arXiv:1412.4419}.

\bibitem{BaYZ}
Balogh F., Yang D., Zhou J., Explicit formula for Witten's $r$-spin partition
 function, in preparation.

\bibitem{BDY1}
Bertola M., Dubrovin B., Yang D., Correlation functions of the {K}d{V}
 hierarchy and applications to intersection numbers over
 {$\overline{\mathcal{M}}_{g,n}$}, \href{https://doi.org/10.1016/j.physd.2016.04.008}{\textit{Phys.~D}} \textbf{327} (2016),
 30--57, \href{http://arxiv.org/abs/1504.06452}{arXiv:1504.06452}.

\bibitem{BDY2}
Bertola M., Dubrovin B., Yang D., Simple {L}ie algebras and topological {ODE}s,
 \href{https://doi.org/10.1093/imrn/rnw285}{\textit{Int. Math. Res. Not.}} \textbf{2018} (2018), 1368--1410,
 \href{http://arxiv.org/abs/1508.03750}{arXiv:1508.03750}.

\bibitem{BDY3}
Bertola M., Dubrovin B., Yang D., Simple Lie algebras, {D}rinfeld--{S}okolov
 hierarchies, and multi-point correlation functions, \href{http://arxiv.org/abs/1610.07534}{arXiv:1610.07534}.

\bibitem{Cafasso}
Cafasso M., Block {T}oeplitz determinants, constrained {KP} and
 {G}elfand--{D}ickey hierarchies, \href{https://doi.org/10.1007/s11040-008-9038-7}{\textit{Math. Phys. Anal. Geom.}} \textbf{11}
 (2008), 11--51, \href{http://arxiv.org/abs/0711.2248}{arXiv:0711.2248}.

\bibitem{CW1}
Cafasso M., Wu C.-Z., Tau functions and the limit of block {T}oeplitz
 determinants, \href{https://doi.org/10.1093/imrn/rnu262}{\textit{Int. Math. Res. Not.}} \textbf{2015} (2015),
 10339--10366, \href{http://arxiv.org/abs/1404.5149}{arXiv:1404.5149}.

\bibitem{CW2}
Cafasso M., Wu C.-Z., Borodin--{O}kounkov formula, string equation and
 topological solutions of {D}rinfeld--{S}okolov hierarchies,
 \href{http://arxiv.org/abs/1505.00556}{arXiv:1505.00556}.

\bibitem{Cartan}
Cartan E., Sur la structure des groupes de transformations finis et continus, Nony et Co, Paris, 1894.

\bibitem{DJKM1}
Date E., Jimbo M., Kashiwara M., Miwa T., Transformation groups for soliton
 equations. {E}uclidean {L}ie algebras and reduction of the {KP} hierarchy,
 \href{https://doi.org/10.2977/prims/1195183297}{\textit{Publ. Res. Inst. Math. Sci.}} \textbf{18} (1982), 1077--1110.

\bibitem{DJKM2}
Date E., Kashiwara M., Miwa T., Transformation groups for soliton equations.
 {II}.~{V}ertex operators and {$\tau $} functions, \href{https://doi.org/10.3792/pjaa.57.387}{\textit{Proc. Japan Acad.
 Ser.~A Math. Sci.}} \textbf{57} (1981), 387--392.

\bibitem{GHM}
de~Groot M.F., Hollowood T.J., Miramontes J.L., Generalized
 {D}rinfel'd--{S}okolov hierarchies, \href{https://doi.org/10.1007/BF02099281}{\textit{Comm. Math. Phys.}} \textbf{145}
 (1992), 57--84.

\bibitem{Dickey}
Dickey L.A., Soliton equations and {H}amiltonian systems, \href{https://doi.org/10.1142/5108}{\textit{Advanced
 Series in Mathematical Physics}}, Vol.~26, 2nd~ed., World Sci. Publ. Co.,
 Inc., River Edge, NJ, 2003.

\bibitem{DS}
Drinfel'd V.G., Sokolov V.V., Lie algebras and equations of {K}orteweg--de
 {V}ries type, \href{https://doi.org/10.1007/BF02105860}{\textit{J.~Math. Sci.}} \textbf{30} (1985), 1975--2036.

\bibitem{Duc}
du~Crest~de Villeneuve A., From the {A}dler--{M}oser polynomials to the
 polynomial tau functions of {K}d{V}, \href{https://doi.org/10.1093/integr/xyx012}{\textit{J.~Integrable Syst.}} \textbf{2}
 (2017), xyx012, 9~pages, \href{http://arxiv.org/abs/1709.05632}{arXiv:1709.05632}.

\bibitem{Du1}
Dubrovin B., Geometry of {$2$}{D} topological field theories, in Integrable
 Systems and Quantum Groups ({M}ontecatini {T}erme, 1993), \href{https://doi.org/10.1007/BFb0094793}{\textit{Lecture
 Notes in Math.}}, Vol.~1620, Springer, Berlin, 1996, 120--348,
 \href{http://arxiv.org/abs/hep-th/9407018}{hep-th/9407018}.

\bibitem{Du2}
Dubrovin B., Gromov--{W}itten invariants and integrable hierarchies of
 topological type, in Topology, Geometry, Integrable Systems, and Mathematical
 Physics, \href{https://doi.org/10.1090/trans2/234/08}{\textit{Amer. Math. Soc. Transl. Ser.~2}}, Vol.~234, Amer. Math.
 Soc., Providence, RI, 2014, 141--171, \href{http://arxiv.org/abs/1312.0799}{arXiv:1312.0799}.

\bibitem{DLZ}
Dubrovin B., Liu S.-Q., Zhang Y., Frobenius manifolds and central invariants for
 the {D}rinfeld--{S}okolov bi{H}amiltonian structures, \href{https://doi.org/10.1016/j.aim.2008.06.009}{\textit{Adv. Math.}}
 \textbf{219} (2008), 780--837, \href{http://arxiv.org/abs/0710.3115}{arXiv:0710.3115}.

\bibitem{DZ-norm}
Dubrovin B., Zhang Y., Normal forms of hierarchies of integrable {PDE}s,
 {F}robenius manifolds and {G}romov--{W}itten invariants,
 \href{http://arxiv.org/abs/math.DG/0108160}{math.DG/0108160}.

\bibitem{Gantmacher}
Gantmacher F.R., The theory of matrices, {V}ols.~1,~2, AMS Chelsea Publishing,
 Providence, RI, 1998.

\bibitem{EH}
Harnad J., Enol'skii V.Z., Schur function expansions of KP {$\tau$}-functions
 associated to algebraic curves, \href{https://doi.org/10.1070/RM2011v066n04ABEH004755}{\textit{Russian Math. Surveys}} \textbf{66}
 (2011), 767--807, \href{http://arxiv.org/abs/1012.3152}{arXiv:1012.3152}.

\bibitem{Hir}
Hirota R., The direct method in soliton theory, \href{https://doi.org/10.1017/CBO9780511543043}{\textit{Cambridge Tracts in
 Mathematics}}, Vol.~155, Cambridge University Press, Cambridge, 2004.

\bibitem{HM}
Hollowood T., Miramontes J.L., Tau-functions and generalized integrable
 hierarchies, \href{https://doi.org/10.1007/BF02098021}{\textit{Comm. Math. Phys.}} \textbf{157} (1993), 99--117.

\bibitem{HMG}
Hollowood T.J., Miramontes J.L., Guill\'{e}n J.S., Additional symmetries of
 generalized integrable hierarchies, \href{https://doi.org/10.1088/0305-4470/27/13/036}{\textit{J.~Phys.~A: Math. Gen.}}
 \textbf{27} (1994), 4629--4644, \href{http://arxiv.org/abs/hep-th/9311067}{hep-th/9311067}.

\bibitem{IZ}
Itzykson C., Zuber J.-B., Combinatorics of the modular group. {II}.~{T}he
 {K}ontsevich integrals, \href{https://doi.org/10.1142/S0217751X92002581}{\textit{Internat.~J. Modern Phys.~A}} \textbf{7}
 (1992), 5661--5705, \href{http://arxiv.org/abs/hep-th/9201001}{hep-th/9201001}.

\bibitem{Kac1978}
Kac V.G., Infinite-dimensional algebras, {D}edekind's {$\eta$}-function,
 classical {M}\"{o}bius function and the very strange formula, \href{https://doi.org/10.1016/0001-8708(78)90033-6}{\textit{Adv. in
 Math.}} \textbf{30} (1978), 85--136.

\bibitem{Kac}
Kac V.G., Infinite-dimensional {L}ie algebras, 3rd~ed., \href{https://doi.org/10.1017/CBO9780511626234}{Cambridge University
 Press}, Cambridge, 1990.

\bibitem{KW}
Kac V.G., Wakimoto M., Exceptional hierarchies of soliton equations, in Theta
 Functions~-- {B}owdoin 1987, {P}art~1 ({B}runswick, {ME}, 1987),
 \href{https://doi.org/10.1090/pspum/049.1}{\textit{Proc. Sympos. Pure Math.}}, Vol.~49, Amer. Math. Soc., Providence, RI,
 1989, 191--237.

\bibitem{Kostant}
Kostant B., The principal three-dimensional subgroup and the {B}etti numbers of
 a complex simple {L}ie group, \href{https://doi.org/10.2307/2372999}{\textit{Amer.~J. Math.}} \textbf{81} (1959),
 973--1032.

\bibitem{Macdonald}
Macdonald I.G., Symmetric functions and {H}all polynomials, 2nd~ed., \textit{Oxford
 Mathematical Monographs}, The Clarendon Press, Oxford University Press, New York, 1995.

\bibitem{NO}
Nimmo J.J.C., Orlov A.Yu., A relationship between rational and multi-soliton
 solutions of the {BKP} hierarchy, \href{https://doi.org/10.1017/S0017089505002363}{\textit{Glasg. Math.~J.}} \textbf{47}
 (2005), 149--168, \href{http://arxiv.org/abs/nlin.SI/0405009}{nlin.SI/0405009}.

\bibitem{Sato}
Sato M., Soliton equations as dynamical systems on an infinite dimensional
 Grassmann manifolds, in Random Systems and Dynamical Systems (Kyoto, 1981),
 \textit{RIMS Kokyuroku}, Vol.~439, Kyoto, 1981, 30--46.

\bibitem{SW}
Segal G., Wilson G., Loop groups and equations of {K}d{V} type, \href{https://doi.org/10.1007/BF02698802}{\textit{Inst.
 Hautes \'{E}tudes Sci. Publ. Math.}} \textbf{61} (1985), 5--65.

\bibitem{S}
Shigyo Y., On the expansion coefficients of tau-function of the {BKP}
 hierarchy, \href{https://doi.org/10.1088/1751-8113/49/29/295201}{\textit{J.~Phys.~A: Math. Theor.}} \textbf{49} (2016), 295201,
 17~pages, \href{http://arxiv.org/abs/1601.02083}{arXiv:1601.02083}.

\bibitem{WilsonAdelic}
Wilson G., Collisions of {C}alogero--{M}oser particles and an adelic
 {G}rassmannian, \href{https://doi.org/10.1007/s002220050237}{\textit{Invent. Math.}} \textbf{133} (1998), 1--41.

\bibitem{Wu}
Wu C.-Z., Tau functions and {V}irasoro symmetries for {D}rinfeld--{S}okolov
 hierarchies, \href{https://doi.org/10.1016/j.aim.2016.10.028}{\textit{Adv. Math.}} \textbf{306} (2017), 603--652,
 \href{http://arxiv.org/abs/1203.5750}{arXiv:1203.5750}.

\bibitem{You}
You Y., Polynomial solutions of the {BKP} hierarchy and projective
 representations of symmetric groups, in Infinite-Dimensional {L}ie Algebras
 and Groups ({L}uminy-{M}arseille, 1988), \href{https://doi.org/10.1142/9789812798343}{\textit{Adv. Ser. Math. Phys.}},
 Vol.~7, World Sci. Publ., Teaneck, NJ, 1989, 449--464.

\bibitem{Zhou}
Zhou J., Explicit formula for {W}itten--{K}ontsevich tau-function,
 \href{http://arxiv.org/abs/1306.5429}{arXiv:1306.5429}.

\bibitem{Zhou2}
Zhou J., Fermionic computations for integrable hierarchies,
 \href{http://arxiv.org/abs/1508.01999}{arXiv:1508.01999}.

\end{thebibliography}
\end{document}